\documentclass[conference]{IEEEtran}
\IEEEoverridecommandlockouts

\usepackage{caption}
\usepackage{subcaption}
\usepackage{cite}
\usepackage{amsmath,amssymb,amsfonts}
\usepackage{algorithmic}
\usepackage[algo2e]{algorithm2e}
\usepackage{algorithm}
\usepackage{graphicx}
\usepackage{siunitx}
\usepackage{textcomp}
\usepackage{xcolor}
\usepackage{siunitx}
\usepackage{balance}
\usepackage{soul}
\usepackage{comment}
\usepackage{epstopdf}

\begin{document}

\title{Phase Optimization and Relay Selection for Joint Relay and IRS-Assisted Communication \\
\thanks{Part of this work was done when Uyoata Uyoata was at Modibbo Adama University, Yola}
}

\author{

\IEEEauthorblockN{Uyoata E. Uyoata\IEEEauthorrefmark{1}\IEEEauthorrefmark{3} Mobayode O. Akinsolu\IEEEauthorrefmark{2}  Enoruwa Obayiuwana  \IEEEauthorrefmark{4}  Abimbola Sangodoyin \IEEEauthorrefmark{5}
 Ramoni Adeogun\IEEEauthorrefmark{1}}

\IEEEauthorblockA{\IEEEauthorrefmark{1} Wireless Communication Network Section, Aalborg University, Aalborg, Denmark.}
\IEEEauthorblockA{\IEEEauthorrefmark{2} Faculty of Arts, Computing, and Engineering, Wrexham University, Wrexham, LL11 2AW, UK}
\IEEEauthorblockA{\IEEEauthorrefmark{4} Department of Electronic and Electrical Engineering, Obafemi Awolowo University, Ife, Nigeria}
\IEEEauthorblockA{\IEEEauthorrefmark{5} School of Computer Science, University of Lincoln, Lincoln, LN6 7TS.}

\IEEEauthorblockA{\IEEEauthorrefmark{3}ueu@es.aau.dk}

}

\maketitle




\begin{abstract}
The use of Intelligent Reflecting Surfaces (IRSs) is considered a potential enabling technology for enhancing the spectral and energy efficiency of beyond 5G communication systems. In this paper, a joint relay and intelligent reflecting surface (IRS)-assisted communication is considered to investigate the gains of optimizing both the phase angles and selection of relays. 
The combination of successive refinement and reinforcement learning is proposed.  
Successive refinement algorithm is used for phase optimization and reinforcement learning is used for relay selection. Experimental results indicate that the proposed approach offers improved achievable rate performance and scales better with number of relays compared to considered benchmark approaches.

\end{abstract}




\maketitle

\section{Introduction}
Recently, intelligent reflecting surface (IRS) has been proposed as a promising enabling technology for beyond fifth-generation (5G) communication systems \cite{9387701}. 
Typically, IRS has a large number of reflecting elements with which the wireless propagation environment can be intelligently tuned to achieve some network design targets, including spectral efficiency and energy efficiency \cite{8910627}. The gains and challenges of integrating IRS into future wireless communication systems such as 6G communication networks have been reported in the literature, for example, see \cite{DiRenzoMarc} and \cite{Liasko}. 
There have also been research efforts to study the performance of systems that combine IRS with existing technologies such as physical layer security \cite{YangHelin}, device-to-device communication \cite{ChenYal}, and relaying \cite{8888223}.   
Due to its similarity with relaying, IRS has been compared to relaying and their co-use has been demonstrated in some works like \cite{9733238} and \cite{9940169}. 
The work in \cite{9108262} derived upper bounds for the ergodic capacity of a hybrid relay and IRS network and proposed a hybrid design which was shown to offer higher available rate compared to using IRS only, especially in low and medium signal-to-noise ratio (SNR) ranges. 

A hybrid multiple-relay and IRS-assisted network is considered in \cite{9344820}  where a joint problem of IRS coefficient optimization and relay selection is presented. The formulated problem is solved using deep reinforcement learning, an approach that results in higher throughput performance when compared to random IRS coefficient and relay selection. In \cite{9508416}, co-optimization of the IRS coefficients and buffer-aided relay selection was studied. Specifically, a multi-agent deep reinforcement learning scheme was proposed to solve the sub-problems of IRS coefficient optimization and relay selection in the presence of an eavesdropper. 
In the literature, where both IRS phase angle optimization and relay selection were studied, a single algorithm was used for both problems as typified in \cite{9344820}. 


In this paper, we present the optimization of IRS phase angles and relay selection. Similar to our proposed approach, the work in \cite{9344820} jointly considers relay selection and IRS reflection coefficient optimization using reinforcement learning for both relay selection and reflection coefficient optimization. However, our proposed approach is distinct from the work in  \cite{9344820} in that we employ a reinforcement learning technique, specifically Q-learning for relay selection, and successive refinement approach for phase angle optimization. In so doing, the gains of both methods are aggregated in a unified framework. Specifically, the following are the contributions of this paper:

\begin{itemize}
\item Decomposing a formulated joint phase angle optimization and relay selection problem into sub-problems to allow the application of separate optimization techniques, namely successive refinement and Q-learning, respectively.

\item Using a search space reduction criterion for populating the Q-learning reward matrix.
\item Experimental validation of proposed approach via simulations.
\end{itemize}

\section{System Model}
\label{system_model}
\begin{figure}[!t]
\includegraphics[width=3.35in]{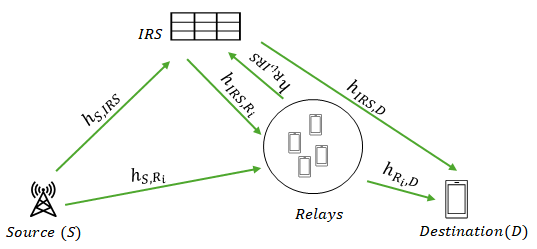}
\centering
\caption{System Model}
\label{F-0}
\end{figure}

The network set-up considered is shown in Fig. \ref{F-0}. It comprises an IRS and a group of relays that assist the communication between a source ($S$), and a destination ($D$). 
The source ($S$), $R$ relays indexed by $\mathcal{R} = \{ 1, 2, 3,..., R\}$, and destination ($D$) are equipped each with a single antenna. 
We assume that the IRS, with $N$ elements (belonging to a set, $\mathcal{N}$), is connected to a controller via which the phase angles of the elements can be intelligently adjusted. For this system model, two time slot communication is considered, whereby in the first communication time slot, the source  transmits towards the relays and the IRS, and the IRS also reflects signals from the source to the relays. In the second communication time slot, a suitably selected relay ($R_{i}$) transmits towards $D$ via the IRS. Similar system model was used in \cite{Joint} where the authors studied joint power allocation, relay selection and beamforming.
The locations of all nodes in the network are described by three dimensional Cartesian coordinate system $(x,y,z)$ and the relays are uniformly distributed within a circular area. These relays are assumed to have decode and forward capabilities. This system model is practical, for example in scenarios where a destination device within a building  can be assisted by nearby IRS  and relay-enabled devices located along a street adjacent the building.

\subsection{Channel Model}
A combination of Line of Sight (LoS) and Non-Line of Sight (NLoS) links is considered. 
Note that a similar LoS and NLoS arrangement is used in \cite{Yildirim}, where the relay to destination link is assumed available. Similar to the work in \cite{DDilin}, we adopt a 3GPP TR 38.901 UMi-Street Canyon path loss model such that:
\begin{equation}
PL_{LOS}= 32.4 + 21log_{10}(d) + 20log_{10}{f_{c}} 
\label{1.1}
\end{equation}
\begin{equation}
PL^{'}_{NLOS}= 
22.4 + 35.5log_{10}(d) + 21.3log_{10}{f_{c}} - 0.3(h_{UT} - 1.5)
\end{equation}

\begin{equation}
PL_{NLOS}= max(PL_{LOS},PL^{'}_{NLOS}),
\label{1.2}
\end{equation}
where $PL_{LOS}$ and $PL_{NLOS}$ are the path loss for the LoS link and path loss for the NLoS link, respectively, $d$ is the separation in metres between any two devices, $f_{c}$ is the carrier frequency in GHz, and $h_{UT}$ is the user terminal height. For the work in this paper, $h_{UT} = 1 m$, which is a typical mobile terminal height. All considered links are modeled using equations (\ref{1.1}), (\ref{1.2}) and Rician small scale fading.  
Based on the foregoing, at the end of the first communication time slot, the $i_{th}$ relay, $R_{i}$ receives a signal given by:
\begin{equation}
y_{R_{i}} = \sqrt{P_{s}}x_{S}(h_{S,R_{i}} +  \textbf{h}_{IRS,R_{i}}^{T} \Phi_{1} \; \textbf{h}_{S,IRS}) + z_{R_{i}},
\end{equation}
where $P_{s}$ is the transmit power of $S$, $x_{S}$ is the information signal having unit power, $h_{S,R_{i}}$ is the channel coefficient of the $S$ to  $R_{i}$ link, $\textbf{h}_{IRS,R_{i}}$ is the channel coefficient of the $IRS$ to $R_{i}$ link and $\textbf{h}_{S,IRS}$ is the channel coefficient of the $S$ to $IRS$ link. Note that the superscript $(.)^T$ refers to the transpose operation. The  Additive White Gaussian Noise at relay $R_{i}$ is given by $z_{R_{i}}$, where $z_{R_{i}}$ has a zero mean and a variance of $\sigma^{2}$ . $\Phi_{1}$ is the diagonal reflection matrix of the IRS in the first communication time slot and it is expressed as ({$\Phi_{1} =  diag(\alpha_{1}^{1}e^{j\phi_{1}^{1}},\alpha_{1}^{2}e^{j\phi_{1}^{2}},...,\alpha_{1}^{N}e^{j\phi_{1}^{N}})$.  $\alpha_{1}^{n}$ is the reflection amplitude coefficient and $\phi_{1}^{n}$ is the phase shift of the $n_{th}$ element of the IRS. It is assumed that $\alpha_{1}^{1} = \alpha_{1}^{2}=,..., \alpha_{1}^{n} = 1$ and $\phi_{1}^{n} \in [0,2\pi]$, since the reflection amplitude coefficient is not the optimization goal. Hence, the rate of link used for communication in the first time slot is given by:

\begin{equation}
C_{t1} = log_{2}\bigg(1 + \dfrac{P_{s}\vert h_{S,R_{i}} +  \textbf{h}_{IRS,R_{i}}^{T} \Phi_{1} \; \textbf{h}_{S,IRS} \vert ^2}{\sigma^2} \bigg),
\label{eqshun}
\end{equation}
where $\vert . \vert$ refers to the absolute operation of a complex number and it applies to all such notations within this paper. Similarly, in the second communication time slot, the signal received by $D$ is given by:

\begin{equation}
y_{D} = \sqrt{P_{R_{i}}}x_{R_{i}}(h_{R_{i}, D} +  \textbf{h}_{IRS,D}^{T} \Phi_{2} \; \textbf{h}_{R_{i},IRS}) + z_{D},
\end{equation}
where $P_{R_{i}}$ is the transmit power of a selected relay, $R_{i}$. $h_{R_{i},D}$, $\textbf{h}_{IRS,D}^{T}$ and $\textbf{h}_{R_{i},IRS}$ are the channel coefficients of $R_{i}$ to $D$, $IRS$ to  $D$ and the $R_{i}$ to $IRS$ links, respectively. The Additive White Gaussian noise at $D$ is given by $z_{D}$, where $z_{D}$ has a zero mean and a variance of $\sigma^{2}$. $\Phi_{2}$ is the diagonal reflection matrix of the IRS in the second communication time slot expressed as $\Phi_{2} =  diag(\alpha_{2}^{1}e^{j\phi_{2}^{1}}, \alpha_{2}^{2}e^{j\phi_{2}^{2}},...,\alpha_{2}^{N}e^{j\phi_{2}^{N}})$.  $\alpha_{2}^{n} $ is the reflection amplitude coefficient and $\phi_{2}^{n}$ is the phase shift of the $n_{th}$ element of the IRS and follow the same assumptions for $\Phi_1$ 

Furthermore, the phase shifts of the IRS elements are assumed to be discrete such that for each element, the phase shift set is given by $\Gamma = [0, \Delta\phi,..., \Delta\phi(K-1)]$, where $\Delta\phi = 2\pi/K$ and $K = 2^{b} $. $b$ is the number of bits used. Thus, the rate of the link  in the second communication time slot is given by:

\begin{equation}
C_{t2} = log_{2}\bigg(1 + \dfrac{P_{R_{i}}\vert h_{R_{i},D} +  \textbf{h}_{IRS,D}^{T} \Phi_{2} \; \textbf{h}_{R_{i},IRS} \vert ^2}{\sigma^2} \bigg).
\label{lo2}
\end{equation}

The rate of the end-to-end link from $S$ to $D$ via $IRS$ and a suitably selected relay $R_{i}$ can be expressed as:
\begin{equation}
C_{S,D} = \dfrac{1}{2}min(C_{t1}, C_{t2}).
\end{equation}

\section{Problem Formulation}
\label{problem_formulation}
 Since this paper aims to co-optimize the phase angles  of the IRS and  the selection of a relay to assist the source - destination communication, the optimization problem is formulated as follows:
\begin{subequations}
\begin{align}
& \textbf{P1:} \: \: \underset{\Phi_{1}, \Phi_{2}, R_{i}}{\text{maximize}} \;C_{S,D}\\  
& \: \: \: \: \: \:\text{subject to:} \notag \\
& \: \: \: \: \: \:C_1 :  \phi_{1}^{n} ,  \phi_{2}^{n}  \in \Gamma \:\: \forall n \in \mathcal{N} &\\
&\: \: \: \: \: \:C_2 : \sum\limits_{i = 1}^{R} x_{i} = 1 &\\
&\: \: \: \: \: \:C_3 :  x_{i} \in \{0,1\}, i =1,2,...,R
\label{eqn_bb_1}
\end{align}
\end{subequations}
In problem $\textbf{P1}$, $C_{1}$, $C_{2}$ and $C_{3}$ are constraints that ensure that the phase angles are members of the discrete phase shift set $\Gamma$ and the number of relay nodes that can assist source - destination communication is limited to $1$. $ x_{i} $ is the relay selection parameter. Maximizing $C_{S,D}$ in \textbf{P1} is equivalent to maximizing $min(C_{t1}, C_{t2})$ . Similarly maximizing  $min(C_{t1}, C_{t2})$ is equivalent to maximizing $min(\gamma_{1}, \gamma_{2})$ where $\gamma_{1}$ is the SNR of the link in the first communication time slot and  $\gamma_{2}$ is the SNR of the link in the second communication time slot. Thus if we drop the logarithmic function in equation}  \ref{eqshun} and \ref{lo2}, \textbf{P1}  can be rewritten using the SNRs of the links in the two communication time slots instead of the end-to-end rate. Hence the problem in \textbf{P1} becomes:
\begin{subequations}
\begin{align}
& \textbf{P2:} \: \: \underset{\Phi_{1}, \Phi_{2}, R_{i}}{\text{maximize}} \;( \text{min}[\gamma_1, \gamma_2])\\  
& \: \: \: \: \: \:\text{subject to:} \notag \\
& \: \: \: \: \: \:C_1 :  \phi_{1}^{n} ,  \phi_{2}^{n}  \in \Gamma \:\: \forall n \in \mathcal{N} &\\
&\: \: \: \: \: \:C_2 : \sum\limits_{i = 1}^{R} x_{i} = 1 &\\
&\: \: \: \: \: \:C_3 :  x_{i} \in \{0,1\}, , i =1,2,...,R,
\label{eqn_bb_2}
\end{align}
\end{subequations} 
where $\gamma_{1} = \dfrac{P_{s}\vert h_{S,R_{i}} +  \textbf{h}_{IRS,R_{i}}^{T} \Phi_{1} \; \textbf{h}_{S,IRS} \vert ^2}{\sigma^2}$
and $\gamma_{2} =  \dfrac{P_{R_{i}}\vert h_{R_{i},D} +  \textbf{h}_{IRS,D}^{T} \Phi_{2} \; \textbf{h}_{R_{i},IRS} \vert ^2}{\sigma^2} $ 
are the SNRs of the first and second communication time slots, respectively. From the expressions for $\gamma_{1}$ and $\gamma_{2}$, it is obvious that for fixed $P_{s}$ and $P_{R_{i}}$, $\gamma_{1}$ depends on $\Phi_{1}$, and $\gamma_{2}$ depends on $\Phi_{2}$, respectively. 
Thus two sub-problems can be derived from $\textbf{P2}$ to obtain first a sub-problem that optimizes the phase angles of the IRS as follows:

\begin{subequations}
\begin{align}
& \textbf{P3:} \: \: \underset{\Phi_{1}, \Phi_{2}}{\text{maximize}} \;(\text{min}[\gamma_1, \gamma_2])\\  
& \: \: \: \: \: \:\text{subject to:} \notag \\
& \: \: \: \: \: \:  \phi_{1}^{n} ,  \phi_{2}^{n}  \in \Gamma \:\: \forall n \in \mathcal{N} & 
\label{eqn_bb_3}
\end{align}
\end{subequations}

and secondly, the relay selection sub-problem:
\begin{subequations}
\begin{align}
& \textbf{P4:} \: \: \underset{R_{i}}{\text{maximize}} \; \lbrace  \text{min}(\gamma_1, \gamma_2)\rbrace\\  
& \: \: \: \: \: \:\text{subject to:} \notag \\
&\: \: \: \: \: \sum\limits_{i = 1}^{R} x_{i} = 1 &\\
&\: \: \: \: \: x_{i} \in \{0,1\}, , i =1,2,...,R
\label{eqn_bb_4}
\end{align}
\end{subequations}

Problem $\textbf{P3}$ is non-convex since the objective function is a non-concave function of  $\Phi_{1}$ and  $\Phi_{2}$ \cite{Yyang}. To ease the difficulty of solving $\textbf{P3}$, the phase angles of the IRS are optimized for each transmission time slot. That is, $\Phi_{1}$ is separately optimized to maximize $\gamma_{1}$ and $\Phi_{2}$ is separately optimized to maximize $\gamma_{2}$. In doing so, the optimization problem in equations \num{12} and \num{13} of \cite{QWu} is similar to the slot by slot optimization of $\Phi_{1}$ and $\Phi_{2}$ to maximize $\gamma_{1}$ and $\gamma_{2}$ respectively. 
Therefore as in \cite{QWu}, successive refinement algorithm can be applied to the separate optimization of $\Phi_{1}$ and $\Phi_{2}$. For example, considering the first communication time slot, the optimization problem, becomes:
\begin{subequations}
\begin{align}
& \textbf{P5:} \: \: \underset{\Phi_{1}}{\text{maximize}} \;\gamma_1\\  
& \: \: \: \: \: \:\text{subject to:} \notag \\
& \: \: \: \: \: \:  \phi_{1}^{n}  \in \Gamma \:\: \forall n \in \mathcal{N} & 
\label{eqn_p5}
\end{align}
\end{subequations}
Similar reasoning holds for the second time slot.
\section{Proposed Approach}
In this section, algorithms to solve the formulated problems are presented. Also, the descriptions of other algorithms that are used as benchmark algorithms for comparisons are also presented in this section.

\subsubsection{Successive Refinement-Based Phase Angle Optimization} 
To solve problem \textbf{P5}, which is non-convex \cite{Yyang}, some expressions in the objective function will need to be simplified. Since the noise term in the expression for $\gamma_{1}$ is a constant, then the numerator can be the focus of the simplification. Let $\theta = \textbf{h}_{IRS,R_{i}} diag(\textbf{h}_{S,IRS})$, $A = \theta^{H}\theta$, $b = \theta^{H}h_{S,R_{i}}$, $v = [e^{j\phi_{1}^{1}}, e^{j\phi_{1}^{2}},...,e^{j\phi_{1}^{N}}]$. Substituting these terms into the numerator of $\gamma_{1} = \dfrac{P_{s}\vert h_{S,R_{i}} +  \textbf{h}_{IRS,R_{i}}^{T} \Phi_{1} \; \textbf{h}_{S,IRS} \vert ^2}{\sigma^2}$ considering that $\frac{P_{s}}{\sigma_{2}}$ is fixed/constant leads to,  
\begin{equation}
\begin{aligned}
\vert h_{S,R_{i}} +  \textbf{h}_{IRS,R_{i}}^{T} \Phi_{1} \; \textbf{h}_{S,IRS} \vert ^2 =  v^{H}Av + 2\mathfrak{Re}\lbrace v^{H}b \rbrace + \\ \vert h_{S,R_{i}}\vert^{2}.
\end{aligned}
\label{eqn12}
\end{equation}
If each phase angle $\phi^{n}_{1}$ in Equation \ref{eqn12} is optimized while keeping the rest of the phase angles constant, the function in Equation \ref{eqn12} becomes linear with respect to the particular phase angle ($\phi^{n}_{1}$) under consideration. That is, if $v_{l}, l\in \mathcal{N} $ is to be optimized, then all other $v_{k},k\neq l, k\in \mathcal{N} $ will be held constant. Equation \ref{eqn12} can be then expressed as:
\begin{equation}
\vert h_{S,R_{i}} +  \textbf{h}_{IRS,R_{i}}^{T} \Phi_{1} \; \textbf{h}_{S,IRS} \vert ^2 = 
2\mathfrak{Re}\lbrace v_{l}w_{l}  \rbrace  +  \sum \limits^{N}_{k\neq l} \sum \limits^{N}_{i\neq l} v_{k}A_{k,i}v_{i} + C
\end{equation}
 where $w_{l} = \sum \limits^{N}_{k\neq l} A_{l,k}v_{k} + b_{k}$ and $C = A_{l,l} + 2\mathfrak{Re} \big\lbrace \sum\limits^{N}_{k\neq l}v_{k}b_{k} \big\rbrace + \vert h_{S,R_{i}}\vert^2$. Therefore, the optimal phase angle can be obtained by optimizing the phase angle that minimizes the difference between each $\phi^{n}_{1}$ and $angle(w_{l})$. The same approach is carried out to obtain the optimal phase angles ($\phi^{n}_{2}$) for the second communication time slot. 

 \begin{algorithm}[!h]
\textbf{Initialize:} $\Phi_{1} = \Phi_{1}^{0}, m = 1$ \\
 \textbf{Input:} $\textbf{h}_{IRS,R_{i}}, P_{s}, h_{S,R_{i}},  \textbf{h}_{S,IRS}, \xi, N$\\
 $C_{t1}^{0} = log_{2}\bigg(1 + \dfrac{P_{s}\vert h_{S,R_{i}} +  \textbf{h}_{IRS,R_{i}}^{T} \Phi_{1} \; \textbf{h}_{S,IRS} \vert ^2}{\sigma^2} \bigg)$\\
\While {$ \vert C_{t1}^{m} - C_{t1}^{m-1}\vert > \xi $}{
 
\For {l = 1 : N} {
$\phi_{1}^{l*} =$  \text{arg  } $min_{\phi_{1} \in \Gamma}$ $\vert \phi_{1} - \angle w_{l} \vert$ \\

}
m = m + 1

 $C_{t1}^{m} = log_{2}\bigg(1 + \dfrac{P_{s}\vert h_{S,R_{i}} +  \textbf{h}_{IRS,R_{i}}^{T} \Phi_{1} \; \textbf{h}_{S,IRS} \vert ^2}{\sigma^2} \bigg)$ \\
}
\Return Optimal phase angles: $\Phi_{1}^{*}$, Optimal achievable rate; $C_{t1}^{*}$
\caption{: Successive Refinement for Phase Optimization \cite{DDilin}}
\end{algorithm}
The algorithm is listed in Algorithm 1. As in \cite{DDilin}, $\xi$ is used as a convergence stop threshold.
By optimizing the angles of the reflecting elements successively (i.e., a single reflecting element at a time), the implementation of Algorithm 1 mirrors a linear search over discrete phase shift levels ($N$ steps in total) to determine the optimal phase shift for the $m_{th}$ reflector. As a result, the worst-case computational complexity of Algorithm 1 can be reasonably deduced to be $\mathcal{O} (n)$ steps, for an input size of $n$, as it is for a typical linear search algorithm \cite{royer2018complexity}. 

With the optimal phase angles of the IRS obtained through Algorithm 1, relay selection is implemented. Using the the optimally selected relay, Algorithm 1 is repeated for the second communication time slot to obtain $\Phi_{2}^{*}$ and $C_{t2}^{*}$ from which the end-to-end link rate is calculated.

We solve the relay selection problem in $\textbf{P}_{4}$ using Q-Learning. We consider the relay selection process as a Markov decision process involving an agent that makes sequential decisions. Details of the Q-Learning based relay selection approach is given in the next subsection.


\subsubsection{Q-Learning Based Relay Selection}
\label{Q_learning}
In Q-learning, an agent learns from its environment and transitions from one state to another by taking certain actions in a way that maximizes the reward of the agent. 
In this work, we assume that the agent is placed at an IRS controller, and the states, $\mathcal{S}$ that make up the learning environment is the relay set. That is, $\mathcal{S} = \lbrace \mathcal{R} = \{ 1, 2, 3,..., R\}\rbrace $. Moreover, the set of possible actions for the agent to choose from is derived from features of the relays. These relay features are specifically the channel gains of the links between the relays and the IRS in the second communication time slot. That is,  the action set, $\mathcal{A} = \lbrace \textbf{h}_{R_{1},IRS}, \textbf{h}_{R_{2},IRS}, \textbf{h}_{R_{3},IRS},...,\textbf{h}_{R_{R},IRS} \rbrace$. The agent transitions between states $s_{t}\in \mathcal{S}$ by taking action $a_{t} \in \mathcal{A}$, where $t$ indicates machine training cycles or episodes, and receives a reward $r(s_{t}, a_{t})$. For each state-action pair, a Q-value, $Q(s_{t}, a_{t})$ is updated and stored in a matrix called the Q-table. 
The Q-table entries are updated based on the well-known Bellman Equation:
\begin{equation}
\begin{aligned}
Q(s_{t}, a_{t})  \leftarrow Q(s_{t}, a_{t}) + \delta\bigg(r(s_{t}, a_{t}) + \eta  \big(max_{a_{t}}Q(s_{t+1},a)\big)\bigg)
\end{aligned}
`\label{Bell}
\end{equation}
where $\delta$ is the algorithm's learning rate, $\delta \in [0\; 1] $ and $\eta$ is the discount factor, $\eta \in [0\; 1]$. 
To transition to a new state, the agent either takes the next action randomly (with a probability $e \in [0, 1]$) and learns from the resulting reward, or takes the next action greedily (with probability, $1 - e$ ) to maximize its immediate reward. 

One of the features of Q-learning is the reward matrix. The elements of the reward matrix are typically the rewards an agent gains for each state transition. In this work, the criteria for populating the reward matrix is based on the channel gain of each relay to IRS link $(\textbf{h}_{R_{i}, IRS})$. The agent transitions between states/relays and observes the reward and this reward is the link gain obtained from moving between relays. 
Using the link gains implies that the values of the link gains should be known. 
Where the IRS is operator-deployed, this information can be made available to the IRS controller. 

In designing the reward matrix ($RW$) for our investigations, for each relay, the gains of the links between each relay and the elements of the IRS were first summed (in other words, the relays to IRS channel gain were obtained) and stored in a diagonal matrix ($DI$). For each relay, the ratio of the relay to IRS link gain, to each entry of $DI$ formed the row entries of $RW$. Where  $RW$  entries were less than $1$, the entries were removed by zeroing. 
In so doing, the search space of the relay matrix is reduced thus reducing complexity. The search space must be reduced because Q-Learning simulates an exhaustive search to select the optimal relay-enabled device \cite{9079922}. The procedure for obtaining the reward matrix is listed in Algorithm 2. The Q-learning-based relay selection algorithm is listed in Algorithm 3. 
\begin{algorithm}
 Input $\textbf{h}_{IRS, R_{i}}$ \\
 DI = $diag( \textbf{h}_{IRS,R_{i}} )$\\
\For {i = 1: $n_R$ }{
 $RW(i,:) = DI./D(i)$\\
 $RW(i,:) < 1 =0$
}
\Return Reward matrix = $RW$
\caption{: Reward Matrix Procedure}
\end{algorithm}
The motivation for using Q-Learning comes from its less cumbersome implementation. Moreover, the breaking of the formulated problem in Section III into phase optimization and relay selection allows the application of a less involved reinforcement learning approach such as Q-Learning.  Since the task representation or initialization of Algorithm 3 is sufficiently adequate in this case, then its worst-case computational complexity of reaching a goal state can be deduced to have a tight upper bound of $\mathcal{O} (n^3)$ action executions for an input size of $n$ \cite{koenig1993complexity}. 
The combination of the successive refinement-based phase angle optimization and the Q-learning-based relay selection forms our proposed Q-learning-based joint IRS and relay-assisted communication (\textit{QL-JIRA}).
An overview of benchmark approaches that are compared with \textit{QL-JIRA} is presented in Section \ref{other_methods}.

\subsubsection{Benchmark Approaches} 
\label{other_methods}
The performance of our proposed \textit{QL-JIRA} is compared with the performance of some benchmark approaches. These benchmark approaches are: (a) Joint Relay and IRS Optimal (\textit{R-IRS Optimal}) scheme, (b) Random Selection  (\textit{RS}), (c) Fixed Phase Algorithm (\textit{FPA}), (d) Random Phase Algorithm (\textit{RPA}) and (e) the \textit{No Relay} approach.

In \textit{R-IRS Optimal}, the optimal phase angles for the IRS are selected using the successive refinement algorithm for the two communication time slots. Furthermore, the relay offering the highest relay to IRS channel gain is selected. This approach and our proposed \textit{QL-JIRA} are similar except that our proposed scheme uses  Q-Learning for relay selection. The \textit{RS} algorithm selects a relay in no particular order, meaning that there is no selection criterion. The random selection only applies to relay selection and not IRS phase angles. In the \textit{RS} algorithm, the phase angles are also optimized using the successive refinement algorithm. 

\begin{algorithm}
\caption{: Q-Learning Based Relay Selection Algorithm}
\textbf{Initialization:} $Q(s_{t}, a_{t}) =  0$, $\eta = 0.8$, $e = 0.7$, $t = 10000 $\\
\textbf{Input} $\textbf{h}_{IRS,R_{i}}$\\
\For {z = 1: t}
{
     $ s_{t} \leftarrow  \text{random choice}(\textbf{h}_{IRS,R_{i}}) $ \\ 
    \eIf {rand $< e $}
        { $a_{t}$ $\leftarrow$  \text{random choice}($\textbf{h}_{IRS,R_{i}}$) }
    {\If{rand $> e $}{
    $a_{t} =  arg\: max(Q(s_{t}, a))$\\}
    }  
    Observe ($s_{t+1}$), Reward ($RW_{t+1}$)\\
    Update Q matrix according to (\ref{Bell})
}
\Return selected relay $R_{i}^{*}$ $\leftarrow$  $Q^{*}$ $(s_{t}, a_{t})$ 
\end{algorithm}

The \textit{FPA} uses optimized IRS phase angles based on the successive refinement algorithm in the first communication time slot and optimal relay selection. Optimal relay selection implies that the relay with maximum relay to IRS channel gain in the second communication time slot is selected. Furthermore, in the second communication time slot, the phase angles are fixed at $\phi_{rpa}$ radians. The \textit{RPA} algorithm differs from the \textit{FPA} in  phase angle selection - whereas in \textit{FPA}, the angles are fixed at $\phi_{rpa}$ radians for the second communication time slot, in \textit{RPA} the angles are randomly selected from the range of allowed angles between $0$ and $ 2\pi$ radians.

\textit{FPA} and \textit{RPA} schemes demonstrate the effect of optimizing the IRS phase angles in only one communication time slot. The \textit{No relay} scheme does not consider the presence of any relay node between the source and the destination. This is similar to the work presented in \cite{DDilin}. In adding this scheme to the list of studied algorithms, the gain of relaying in IRS systems is further highlighted.

Note that for our proposed QL-JIRA scheme, the channel model is such that there is a dominant LoS path between the $S$ and IRS, between IRS and $D$ and between the IRS and the relays. The link between $S$ and $D$ is assumed to be obstructed and the links between the relays and $D$ are assumed to be obstructed. The links between the source, $S$ and the relays are considered obstructed as well.
\section{Discussion of Results}

\begin{table}
 \caption{Simulation Parameters}
\label{table_Parameter}
\centering
\begin{tabular}{|l|l|}
\hline
\textbf{Parameter}  & \textbf{Value}\\
\hline
Carrier frequency ($f_{c}$) & 24.2 GHz \\

\hline
Number of discrete levels & 16 \\

\hline
Number of intelligent reflecting surfaces & 256 \\

\hline
Number of relays & 5- 30\\
\hline
Noise power ($\sigma^{2}$)& -60 dBm\\
\hline
Q-Learning Discount factor($\eta$)& 0.8\\
\hline
e-greedy factor ($e$)&0.7\\
\hline
$\phi_{rpa}$ &2.1 radians\\
\hline
\end{tabular}
\end{table}

In this section, simulation results are presented and discussed. The simulation parameters used for obtaining these results are given in Table \ref{table_Parameter}. 
\begin{figure}[!h]
\includegraphics[width=2.75in]%
{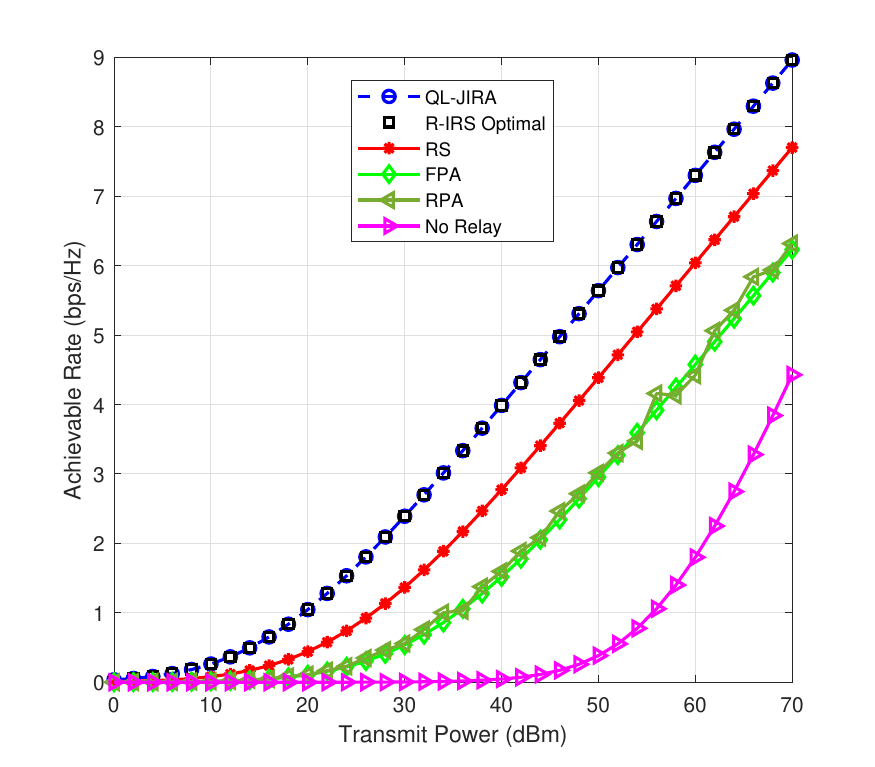}
\centering
\caption{Achievable Rate (bps/Hz) vs. Transmit Power (dBm)} 
\label{F1}
\end{figure}

\begin{figure*}[!h]
    \begin{minipage}[t]{.33\linewidth}
    \includegraphics[width=\linewidth]
    {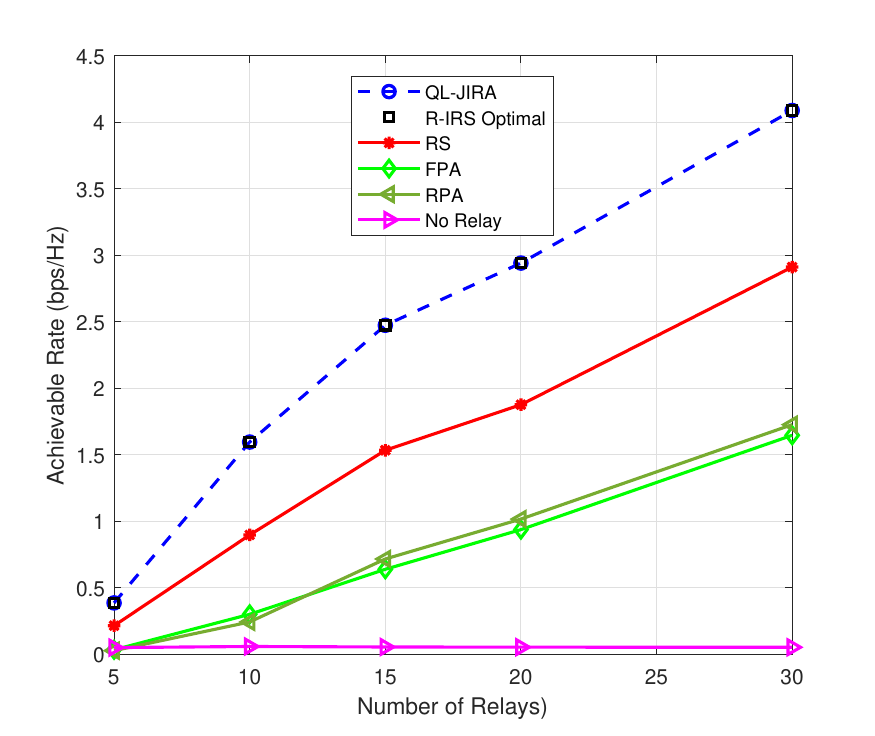}
    \caption{$C_{S,D}$ vs. Number of relays}%
    \label{F0}
  \end{minipage}%
  \begin{minipage}[t]{.33\linewidth}
    \includegraphics[width=\linewidth]{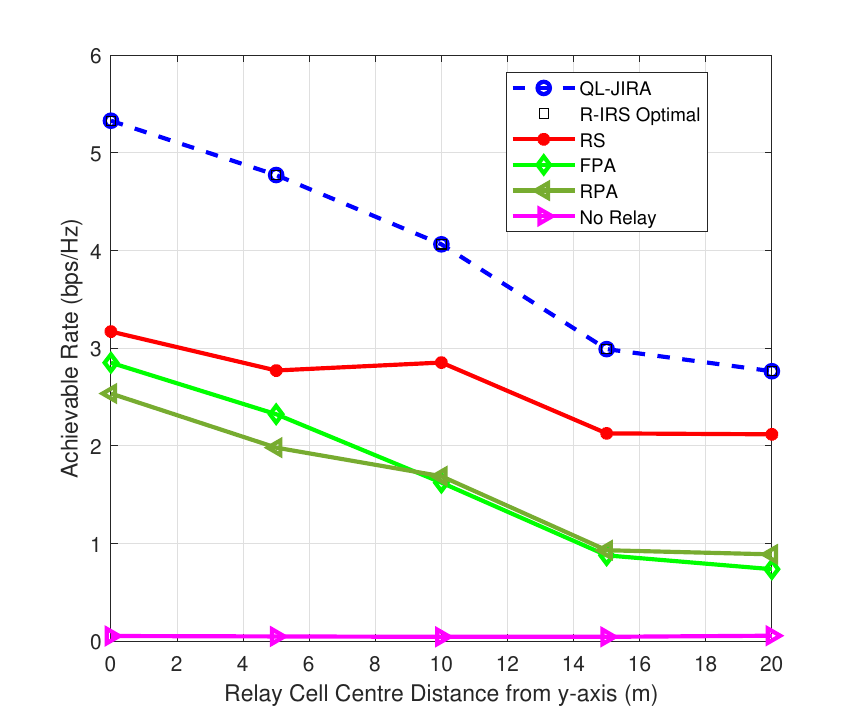}
    \caption{$C_{S,D}$ vs. Relay Cell Centre Distance along the y-axis}%
    \label{F2}
  \end{minipage}\hfil
  \begin{minipage}[t]{.33\linewidth}
    \includegraphics[width=\linewidth]
    {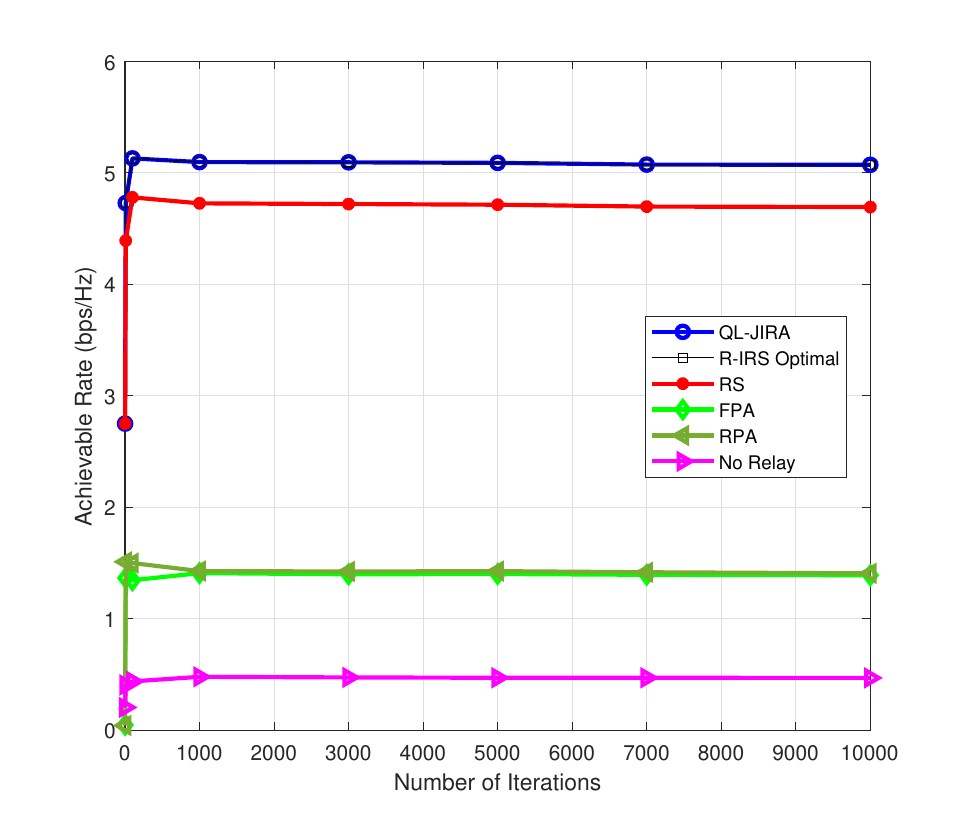}
    \caption{Algorithm Convergence}
    \label{F4}
  \end{minipage}%

\end{figure*}
\subsection{Achievable Rate vs. Transmit Power}
In Fig. \ref{F1}, the achievable rate of the setup is plotted against the transmit power of the transmitting devices in the network. The transmit power is varied from \SI{0}{dBm} to \SI{70}{dBm} and the achievable link rate is observed. In Fig. \ref{F1}, an upward trend can be observed as the transmit power is increased; this trend is observed for all the studied approaches.

The proposed \textit{QL-JIRA} scheme outperforms \textit{RS, FPA, RPA,} and \textit{No-Relay} methods. For example at \SI{40}{dBm}, \textit{QL-JIRA} offers an achievable link rate of \SI{4}{bps/Hz} in comparison with achievable rates of \SI{3}{bps/Hz} and \SI{1.5}{bps/Hz} from \textit{RS} and \textit{FPA}, respectively. Fig. \ref{F1} shows that combining relays and IRS offers significant gains in comparison with having only IRS-assisted setups. 
It can also be deduced from Fig. \ref{F1} that only focusing on optimizing the relay selection is not sufficient for a joint relay and IRS-assisted communication system. This is because the phase angles of the IRS have to be carefully selected as well, as can be seen in the same Fig. \ref{F1}, where a random relay selection scheme with an optimized IRS phase angle  (i.e., \textit{RS} approach) results in a better achievable rate when compared with a scenario having maximum link gain relay selection and a fixed IRS phase angle/randomly selected IRS phase angle (i.e., \textit{FPA} or \textit{RPA} respectively). This goes to show that in joint relay and IRS-assisted communication systems, the IRS phase angle selection could constitute a bottleneck. For the results in Fig. \ref{F1}, the source, IRS, and the destination are located at (\num{20},-\num{10},\num{2}), (\num{0},\num{0},\num{1}), and (\num{20},\num{20},\num{1}), respectively, while the relays are uniformly distributed within a circle of radius \SI{10}{m} with center at (\num{10} \num{10}, \num{0}). 

\subsection{Achievable Rate vs. Number of Relays}
The effect of varying the number of relays on the achievable rate of the considered setup for a fixed  transmit power (of \SI{40}{dBm}) is shown in Figure \ref{F0}. A joint IRS and relay aided communication setup benefits from multiple relays because relay selection can be better optimized due to increased likelihood of having relays with better link to the IRS. Our proposed approach scales well with increased number of relays unlike the \textit{No-Relay} approach.

\subsection{Achievable Rate vs. Distance of Relay Cell Centre along the y-axis}

Fig.~\ref{F2} shows how the achievable rate varies when the relay-cell center is moved along the y-axis (i.e. the relay centre is moved from (\num{10},\num{0},\num{0}) to (\num{10},\num{20},\num{0}). In Fig.~\ref{F2}, the source, destination and IRS locations are fixed and the radius of the cluster of relays is fixed at \SI{10}{m} while the transmit power of the nodes is fixed at \SI{40}{ dBm}. It is easy to see that for \textit{No-Relay}, variations in the position of the relay cell center from the y-axis does not have any impact on the achievable rate, whereas, for other methods, there is a drop in the achievable rate as the relay cell center moves away from the y-axis. 

From the foregoing, it is evident that for setups like the one modeled in this paper, positioning a group of relays near IRS  can guarantee better achievable rates. For example, when the relay cell center is at (\num{10},\num{0},\num{0}), all the considered approaches except the  \textit{No Relay} approach achieves their highest achievable rate. This implies that \textit{QL-JIRA} performs better in distance regimes that are closer to the IRS and this is because relay to destination link is considered obstructed hence proximity to the IRS is more beneficial than proximity to destination. Generally, the performance of the \textit{RS} scheme has been closest to our proposed \textit{QL-JIRA} approach , therefore dynamic scheme selection algorithms that can switch between \textit{QL-JIRA} and \textit{RS} could be useful and practical for varying relay cell center distances. Furthermore, the convergence of the studied approaches are shown in  Fig. \ref{F4} wherein beyond \num{2000} iterations, the achievable rate of the \textit{QL-JIRA} converges to \SI{5}{bps/Hz} for fixed source, destination and IRS locations and at transmit power of \SI{30}{dBm}. And this indicates that our proposed \textit{QL-JIRA} converges within similar iteration vicinity of the compared approaches.

\balance


\section{Conclusion}

In this paper, combining successive refinement-based phase optimization and Q-learning-based relay selection have been proposed as a method for joint IRS and relay-assisted communication. The proposed scheme was shown to offer improved achievable rate over some benchmark approaches. It also scaled with the number of relays.
The simulation results also showed that the selection of the phase angles constitutes a bottleneck for joint IRS and relay communication setups. The experiments further showed that at a fixed transmit power, the relay cell centre needs to be located near the IRS. 

\bibliographystyle{IEEEtran}


\bibliography{irs_arxiv}

\end{document}